\documentclass[fleqn,twoside,twocolumn,nofootinbib]{revtex4} 
\usepackage{ujp} 
\begin{document}
\title[MAGNETOOPTICAL STUDY OF SPIN-ORIENTATION PHASE TRANSITION]
{MAGNETOOPTICAL STUDY OF SPIN-ORIENTATION \\PHASE TRANSITION IN
NdFe\boldmath$_{3}$(BO$_{3}$)$_{4}$ SINGLE CRYSTAL}
\author{V.A.~Bedarev}
\affiliation{B.I.~Verkin Institute for Low Temperature Physics and
Engineering,\\
Nat. Acad. of Sci. of Ukraine}
\address{47, Lenin Av., Kharkiv 61103, Ukraine}
\email{bedarev@ilt.kharkov.ua}
\author{M.I.~Pashchenko}
\affiliation{B.I.~Verkin Institute for Low Temperature Physics and
Engineering,\\
Nat. Acad. of Sci. of Ukraine}
\address{47, Lenin Av., Kharkiv 61103, Ukraine}
\author{D.N.~Merenkov}
\affiliation{B.I.~Verkin Institute for Low Temperature Physics and
Engineering,\\
Nat. Acad. of Sci. of Ukraine}
\address{47, Lenin Av., Kharkiv 61103, Ukraine}
\author{L.N.~Bezmaternykh}
\affiliation{L.V. Kirenskii Institute of Physics, Siberian Branch of the
Russian Academy of Sciences}
\address{Krasnoyarsk 660036, Russia}
\author{V.L.~Temerov}
\affiliation{L.V. Kirenskii Institute of Physics, Siberian Branch of the
Russian Academy of Sciences}
\address{Krasnoyarsk 660036, Russia}

\udk{537.622.5} \pacs{78.20.Ls; 75.50.Ee} \razd{\secvii}

\setcounter{page}{648}%
\maketitle



\begin{abstract}
The magnetic field dependences of birefringence in a
NdFe$_{3}$(BO$_{3}$)$_{4}$ single crystal have been measured in the
case where the direction of light propagation coincides with the
trigonal crystal axis $C_{3}$ ($\mathbf{k}\Vert C_{3}$), and the
external magnetic field is oriented along the second-order axis
$C_{2}$ ($\mathbf{H}\Vert C_{2}$). In the temperature range, in
which an incommensurate phase exists with the formation of a
long-period antiferromagnetic helix, the strongly pronounced jumps
in the field dependence of birefringence are revealed and identified
as a first-order spin-orientation phase transition. The phase
transition was accompanied by a hysteresis in the field dependences
of birefringence. The \textit{H--T} phase diagram for a
NdFe$_{3}$(BO$_{3}$)$_{4}$ single crystal has been plotted in the
case where the magnetic field is oriented along the crystal axis
\textit{C}$_{2}$ ($\mathbf{H}\Vert C_{2}$).
\end{abstract}

\section{Introduction}

A NdFe$_{3}$(BO$_{3}$)$_{4}$ single crystal with the giant
magnetoelectric effect \cite{1} is one of the most intensively
studied ferroborates. Unlike the majority of similar compounds, the
structural phase transition in this crystal was not observed
\cite{2}. The crystal symmetry R32 stays down to helium
temperatures. At a temperature of about 30$~\mathrm{K}$, there
appear the simultaneous magnetic and electric orderings in the
crystal \cite{3}. The NdFe$_{3}$(BO$_{3}$)$_{4}$ single crystal is
antiferromagnetically ordered, and the magnetic moments of Fe and Nd
atoms are parallel to each other in the plane perpendicular to the
trigonal axis $C_{3}$, being oriented oppositely to the magnetic
moments in the neighbor plane \cite{4}. As a result, there appear
three types of antiferromagnetic domains. The antiferromagnetic
vector $\mathbf{l}$ in each of them is oriented along the
corresponding axis of the second order. A further reduction of the
crystal temperature down to 13.5$~\mathrm{K}$ results in a
first-order spin-orientation phase transition from the commensurate
phase into the incommensurate one, the latter structurally being a
long-period antiferromagnetic helix \cite{4}.

It is known that a spin-orientation phase transition takes place in
NdFe$_{3}$(BO$_{3}$)$_{4}$ crystals under the action of a magnetic
field applied along the second-order axis \cite{1,5}. The hysteresis
testifying to a  first-order spin-orientation phase transition  can
be clearly observed in the field dependences of the magnetization.
However, this phase transition is strongly expended over the
magnetic field. This effect may probably be a result of a polydomain
structure of specimens, as well as owing to elastic stresses and
defects in them. Therefore, it is rather difficult to determine the
magnitude of phase transition field. In such cases, the
magnetooptical method often turns out useful. It allows the small
non-strained sections of single-crystalline plates a few tens of
microns in thickness to be selected and studied. In works
\cite{6,7}, this method was applied for the first time to research
the spin-orientation phase transition induced by the magnetic field
in a NdFe$_{3}$(BO$_{3}$)$_{4}$ crystal. The results obtained were
confirmed by the data of resonance \cite{8} and acoustic \cite{9}
studies of the crystal concerned.

In this work, we report the results of our magnetooptical researches of
the spin-orientation phase transition in NdFe$_{3}$(BO$_{3}$)$_{4}$
ferroborate. The work is based on the analysis of the data obtained from the
field dependences of magnetic birefringence in this single crystal.

\section{Experimental Technique}

We studied a triangular single-crystalline
NdFe$_{3}$(BO$_{3}$)$_{4}$ plate, each side of which was equal to
1.5~mm, and the thickness was 70~$\mu \mathrm{m}$. The developed
surface of the crystal was perpendicular to the trigonal $C_{3}$
axis. In order to remove elastic stresses which appeared after the
mechanical treatment, the specimen was annealed at a temperature of
700$~^{\circ}\mathrm{C}$ for 10~h. The crystal to study was inserted
into a solenoid located in an optical helium cryostat. The magnetic
field was directed along the $C_{2}$ axis. The specimen temperature
was measured with a carbon thermometer.

The light birefringence $\Delta n$ is related to the phase difference
$\delta ,$ which arises between two characteristic linearly polarized waves
at the output from a crystal of thickness $t$, as follows:
\begin{equation}
\Delta n=\delta\lambda/2\pi t,
\end{equation}
where $\lambda$ is the light wavelength. In measurements of
$\delta$, we used a quarter-wave plate. The diagram of the
experimental setup for $\delta $-measurements is depicted in Fig.~1.
Light produced by incandescent lamp (\textit{1}) passed through
thermal filter (\textit{2}) and afterward through interference
filter (\textit{3}) with the transmittance maximum at the wavelength
$\lambda=633$\textrm{~nm} and a transmission band of 11\textrm{~nm}.
Further, light passed through polarizer (\textit{5}) and was focused
on specimen (\textit{6}) making use of lens (\textit{4}). The
aperture diameter in diaphragm (\textit{9}) located in the image
plane of lens (\textit{8}) determined the dimension of the specimen
area to study. The diameter of the examined region was about
100~$\mu$m. Elliptically polarized light, which came out of the
crystal, is transformed by  $\lambda/4$ plate (\textit{10}) in
linearly polarized one.  The polarization plane was rotated at that
by the angle $\delta/2$ with respect to the polarization plane of
light incident onto the crystal. To measure the angle $\delta/2$,
the modulation technique was applied with the modulation of the
light polarization plane (modulator (\textit{11})) and the
synchronous detection (amplifier (\textit{14})). The output signal
of the amplifier was applied to an input of personal computer
(\textit{15}).

\section{Experimental Results and Their Discussion}

In the paramagnetic range, the optical indicatrix of
NdFe$_{3}$(BO$_{3}$)$_{4}$ crystal, characterized by the point
crystal group 32, is an ellipsoid of rotation around the
$C_{3}$-axis. The occurrence of a magnetic ordering or the
application of a magnetic field can reduce the optical class of the
crystal to the biaxial one. Let us expand the symmetric part of the
dielectric permittivity $^{S}\varepsilon_{ij}$ in a series in the
magnetic field $H$ and confine the expansion to the terms quadratic
in $H$:
\begin{equation}
^{S}\varepsilon_{ij}=^{S}\varepsilon_{ij}^{0}+\Delta^{S}\varepsilon
_{ij}+q_{ijl}H_{l}+\beta_{ijlk}H_{l}H_{k}.
\end{equation}

\begin{figure}
\includegraphics[width=\column]{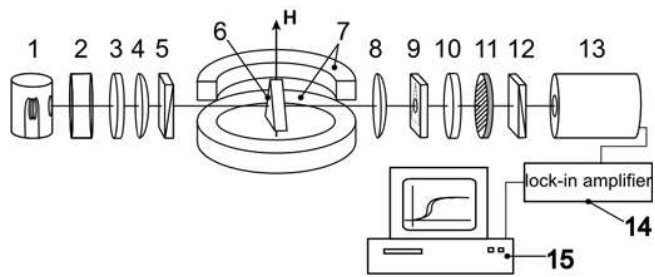}
\vskip-3mm\caption{Diagram of experimental setup:
(\textit{1})~incandescent lamp, (\textit{2})~thermal filter,
(\textit{3})~interference filter, (\textit{4}) and
(\textit{8})~lenses, (\textit{5})~polarizer, (\textit{6})~specimen,
(7)~superconducting solenoid, (\textit{9})~diaphragm, (\textit{10})
$~\lambda /4$-plate, (\textit{11})~modulator,
(\textit{12})~analyzer, (\textit{13})~photoelectronic multiplier,
(\textit{14})~amplifier, (\textit{15})~personal computer  }
\end{figure}

\noindent  Here, $^{S}\varepsilon_{ij}$ is the symmetric part of the
tensor of dielectric permittivity of the crystal in the paramagnetic
range, and $\Delta ^{S}\varepsilon_{ij}$ is a variation of the
symmetric part of the dielectric permittivity tensor associated with
a magnetic and electric ordering. The tensor $q_{ijl}$ describes the
birefringence, which is proportional to the magnetic field strength
and changes its sign, if the direction of the field is opposite. The
tensor $q_{ijl}$ can possess nonzero components only in magnetically
ordered crystals \cite{10}. The tensor $\beta_{ijlk}$ defines the
birefringence proportional to $H_{l}H_{k}$. This tensor is symmetric
with respect to the $(i,j)$ and $(l,k)$ index pairs and is
determined by the point crystal group 32. Thus, the contribution to
the birefringence, which is quadratic in $H$, does not depend on the
magnetic symmetry of the crystal, being determined only by the
crystal symmetry. In the coordinate system with $z\parallel C_{3}$
and $x\parallel C_{2}$, the matrix of $\beta_{ijlk}$-coefficients
looks like
\begin{equation}
\left[ \begin{matrix}
\beta_{11} & \beta_{12} & \beta_{13} & \beta_{14} & 0 & 0 \\
\beta_{12} & \beta_{11} & \beta_{13} & -\beta_{14} & 0 & 0 \\
\beta_{31} & \beta_{31} & \beta_{33} & 0 & 0 & 0 \\
\beta_{41} & -\beta_{41} & 0 & \beta_{44} & 0 & 0 \\
0 & 0 & 0 & 0 & \beta_{44} & 2\beta_{41} \\
0 & 0 & 0 & 0 & \beta_{14} & \beta_{11}-\beta_{12} \\ &  &  &  &  &
\end{matrix}
\right].
\end{equation}
To write down this tensor, we used the standard rules of index notation. In
the case $\mathbf{H}\parallel x$, the additives $\beta_{xxxx}H_{x}^{2}$ and
$\beta_{xxyy}H_{x}^{2}$ to the tensor $\varepsilon_{ij}$ appear in the
magnetic field, in accordance with matrix (3). As a result, the
cross-section of the optical indicatrix of the crystal in a plane
perpendicular to the $C_{3}$-axis is an ellipse, and the crystal becomes
optically biaxial. One of the principal axes of this ellipse is parallel to the
$x$-axis.

Our researches revealed no spontaneous magnetic birefringence defined by
the tensor $\Delta^{S}\varepsilon_{ij}$ and associated with a transition
into the antiferromagnetic state. A spontaneous magnetic birefringence was
also not found, when the crystal transformed into the incommensurate
magnetic phase. The linear magnetic birefringence was observed only when a
magnetic field was applied.

\begin{figure}
\includegraphics[width=6.7cm]{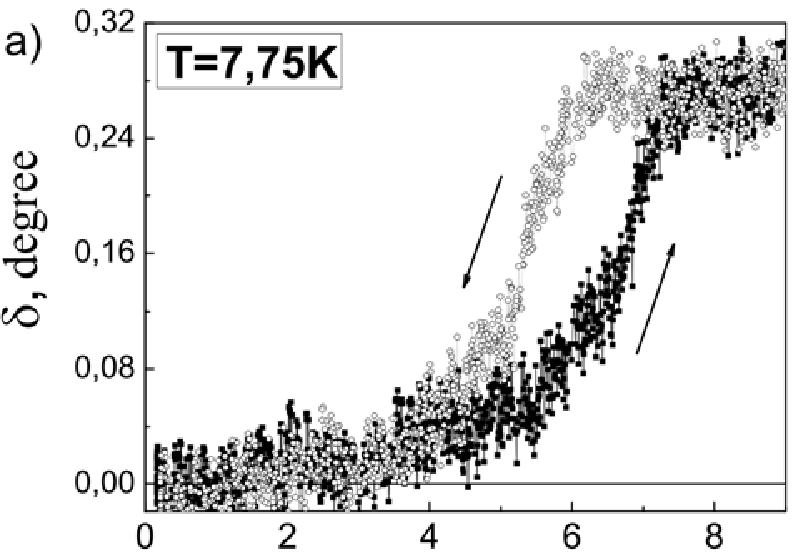}\\ [3mm]
\includegraphics[width=6.7cm]{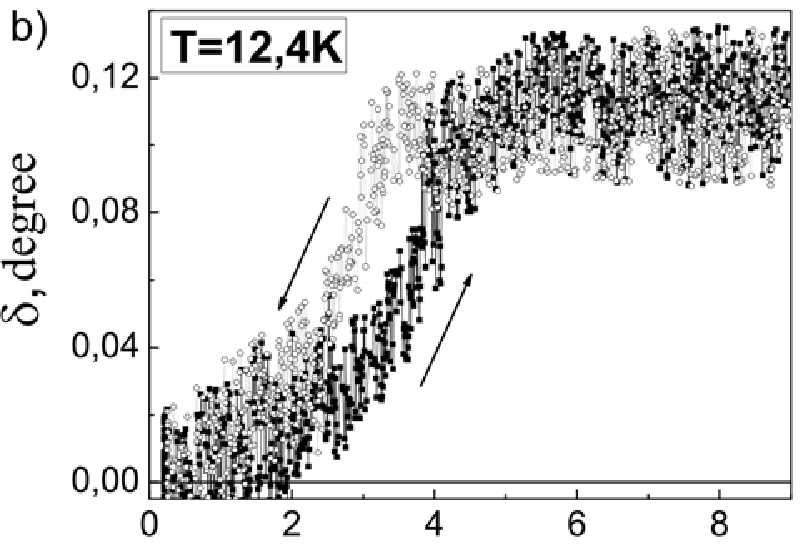}\\ [3mm]
\includegraphics[width=6.7cm]{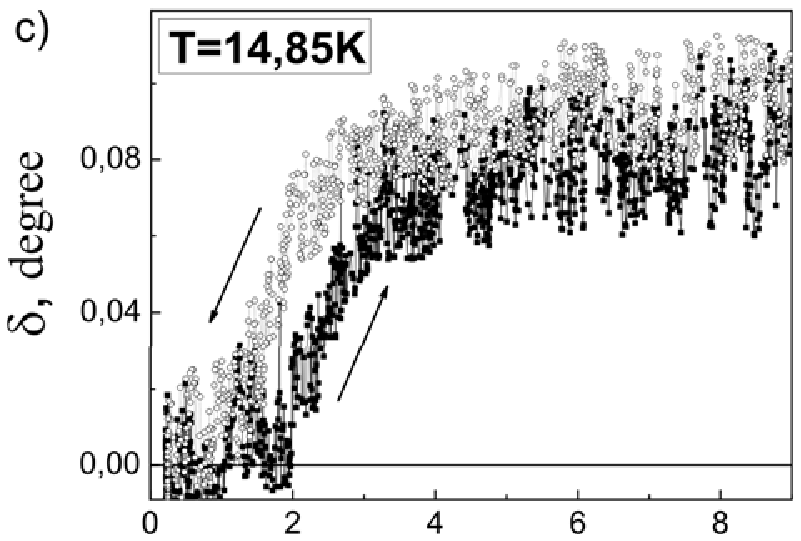}\\ [3mm]
\includegraphics[width=6.7cm]{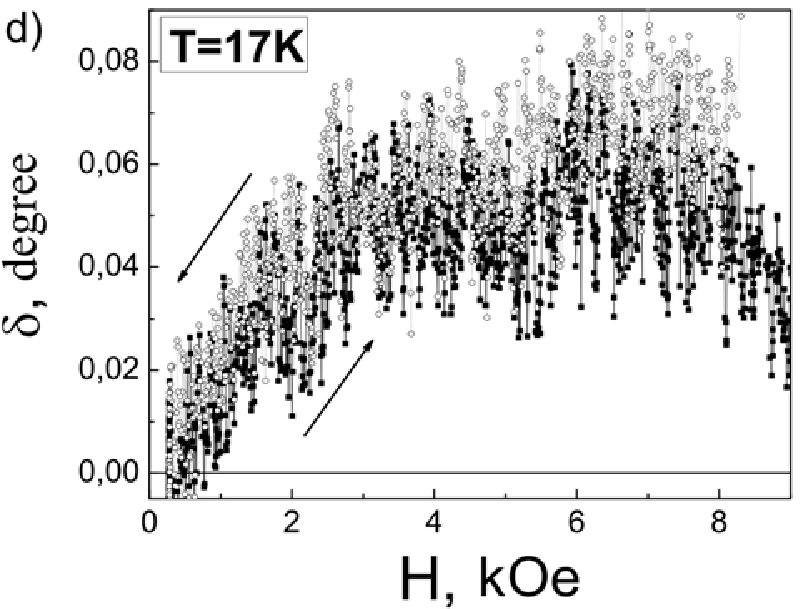}
\vskip-2mm\caption{Field dependences of the light birefringence in a
NdFe$_{3}$(BO$_{3}$)$_{4}$ single crystal measured in the geometry
$\mathbf{k}\parallel C_{3}$ and $\mathbf{H}\parallel C_{2}$ at
various temperatures $T=7.75$ ($a$), 12.4 ($b$), 14.85 ($c$), and
17~K ($d$)  }
\end{figure}

\begin{figure}
\includegraphics[width=6.5cm]{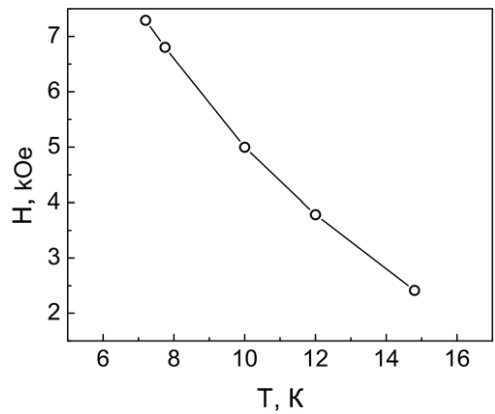}
\vskip-3mm\caption{Magnetic phase diagram \textit{H--T} of a
NdFe$_{3}$(BO$_{3}$)$_{4}$ single crystal for the magnetic field
orientation $\mathbf{H}\parallel C_{2}$  }
\end{figure}

In Fig.~2, the experimental field dependences of the linear magnetic
birefringence of light, $\delta(H)$, measured at various
temperatures from 7.5 to 17~K in the geometry $\mathbf{H}\parallel
x$ are shown. If the direction of the magnetic field is inverted,
the dependence $\delta(H)$ does not change its sign. The
birefringence magnitude is identical for magnetic fields $H$ equal
by the value, but opposite by the direction. Therefore, the observed
magnetic birefringence quadratically depends on the magnetic field,
and the linear magnetic birefringence described by the tensor
$q_{ijl} $ makes no contribution to those dependences. In contrast
to the field dependences of the magnetization $M(H)$ \cite{1}, our
experimental curves demonstrate a pronounced jump $\delta(H)$, which
corresponds to the first-order spin-orientation phase transition.
Similarly to the dependences $M(H)$ \cite{1}, our experimental
curves have a hysteresis in the phase transition interval at low
temperatures. In fields that exceed the transition fields, the
birefringence weakly depends on $H$. The magnitude of $\delta(H)$
and the hysteresis width decrease with the temperature growth. At a
temperature of about 17~K, the first-order phase transition induced
by the magnetic field is not observed any more.\looseness=1

The obtained experimental data allowed us to determine the phase transition
fields at various temperatures. The phase transition field was determined as
follows. First, at a fixed temperature, we determined the field values
corresponding to the middle points in the intervals, where the dependence
$\delta(H)$ changed drastically, at both the increasing and decreasing
external magnetic fields. The phase transition field was calculated as an
average value for those two quantities. The obtained data were used to plot
the phase diagram, which is exhibited in Fig.~3.

It is of interest that the first-order phase transition induced by a
magnetic field was observed only in the incommensurate phase, in
which the long-period antiferromagnetic helix is formed. At
temperatures above 15$~\mathrm{K}$, one could expect that a
spin-flop phase transition would have taken place in the
antiferromagnetic commensurate phase in the case
$\mathbf{H}\parallel C_{2}$. However, it was not observed in our
experiment, although this phase transition was observed in another
easy-plane antiferromagnetic ferroborate, GdFe$_{3}$(BO$_{3}$)$_{4}$
\cite{11,12}. The absence of spin-flop phase transition in our case
can be associated with a deviation of the magnetic field orientation
from the axis of easy magnetization by an angle that exceeds a
certain critical value $\psi_{c}$. The spin-flop transition is known
to disappear in this case, and the rotation of the
antiferromagnetism vector occurs smoothly \cite{13,14,15}.
Neglecting the influence of demagnetizing fields, it is possible to
assert \cite{13,14,15} that the $\psi_{c}$-value is of the order of
the ratio $H_{a}/H_{e}$ between the effective fields of magnetic
anisotropy and exchange antiferromagnetic interaction in the
crystal. Ferroborate NdFe$_{3}$(BO$_{3}$)$_{4}$ is known to have the
high effective exchange field $H_{e}=580~\mathrm{kOe}$ \cite{5} and
the effective field of magnetic anisotropy $H_{a}$, which is very
insignificant in the basic plane (it amounts to about 60~Oe
according to resonance researches \cite{8} and to 12~Oe according to
magnetic ones \cite{5}). The estimation of $\psi_{c}$ gives a
magnitude of 0.3$^{\prime}$. Such a small value of the critical
angle may probably result in that we did not observe the first-order
phase transition in the antiferromagnetic phase of
NdFe$_{3}$(BO$_{3}$)$_{4}$ crystal. On the other hand, the reason
for why the spin-flop transition is distinctly observed in
ferroborate GdFe$_{3}$(BO$_{3}$)$_{4}$, for which the ratio
$H_{a}/H_{e}$ is rather close to ours, may be the influence of
demagnetizing fields in bulk specimens, which favor the increase of
$\psi_{c}$ \cite{13,14,15}. We studied an NdFe$_{3}$(BO$_{3}$)$_{4}$
plate about 70~$\mu\mathrm{m}$ in thickness, and the magnetic field
was oriented in the plate plane. Therefore, the arising
demagnetizing fields were much lower in our case than the
demagnetizing fields appearing in a bulk GdFe$_{3}$(BO$_{3}$)$_{4}$
specimen \cite{11,12}. Hence, they do not affect the critical angle
in our experiment.\looseness=1

\section{Conclusions}

To summarize, while studying the field dependences of the magnetic
birefringence in a NdFe$_{3}$(BO$_{3}$)$_{4}$ single crystal, we
found that the magnetic field $\mathbf{H}\parallel C_{2}$ induces a
first-order spin-orientation phase transition  only in the
temperature range, where the incommensurate phase exists and the
antiferromagnetic helix is realized. In the temperature range, where
the commensurate antiferromagnetic phase exists, the first-order
phase transition in the magnetic field $\mathbf{H}\Vert C_{2}$ is
not observed. We associate this fact with the smallness of the
critical angle $\psi_{c}$ for the magnetic field deviation from the
axis of easy magnetization. The field dependences of the magnetic
birefringence were used to determine the fields of the  first-order
spin-orientation phase transition at various temperatures. The
corresponding magnetic phase diagram \textit{H--T} was plotted.

\vskip3mm The authors express their gratitude to S.L.~Gnat\-chen\-ko
for the useful discussion and valuable remarks.

\rezume {МАГНІТООПТИЧНЕ ДОСЛІДЖЕННЯ ІНДУКОВАНОГО\\ МАГНІТНИМ ПОЛЕМ
СПІН-ОРІЄНТАЦІЙНОГО\\ ФАЗОВОГО ПЕРЕХОДУ У МОНОКРИСТАЛІ\\
NdFe$_{3}$(BO$_{3}$)$_{4}$}{В.А.~Бєдарєв, М.І.~Пащенко,
Д.М.~Мєрєнков, \\Л.М.~Безматерних, В.Л.~Тємєров} {Проведено
дослідження польових залежностей магнітного двозаломлення світла у
монокристалі NdFe$_{3}$(BO$_{3}$)$_{4}$ у випадку, коли напрямок
поширення світла збігається з тригональною віссю кристала $C_{3}$
(${\bf k}\|C_{3}$), а зовнішнє магнітне поле орієнтоване вздовж осі
другого порядку $C_{2}$ (${\bf H}\|C_{2}$). У температурному
проміжку існування неспіврозмірної фази з утворенням
довгоперіодичної антиферомагнітної спіралі виявлено чітко виражені
скачки на польових залежностях двозаломлення, які ідентифіковані як
спін-орієнтаційний фазовий перехід першого роду. Фазовий перехід
супроводжувався гістерезісом на польових залежностях двозаломлення.
Побудовано магнітну фазову \textit{H--T} діаграму
NdFe$_{3}$(BO$_{3}$)$_{4}$ для орієнтації магнітного поля вздовж осі
другого порядку кристала ${\bf H}\|C_{2}$.}

\end{document}